\documentclass[preprint,amsfont,amsmath]{revtex4-1}
\usepackage{graphicx}

\newcommand{\braket}[2]{\langle #1 \mid #2 \rangle}
\newcommand{\etal}[0]{\textit{et al.}}

\begin{document}

\title{Mass dependence of short-range correlations in nuclei and the EMC effect}

\author{Maarten Vanhalst}
\email{Maarten.Vanhalst@UGent.be}
\author{Jan Ryckebusch}
\email{Jan.Ryckebusch@UGent.be}
\author{Wim Cosyn}
\email{Wim.Cosyn@UGent.be}

\affiliation{Department of Physics and Astronomy,\\
 Ghent University, Proeftuinstraat 86, B-9000 Gent, Belgium}

%\begin{start}

\begin{abstract}

  An approximate method to quantify the mass dependence of the number
  of two-nucleon (2N) short-range correlations (SRC) in nuclei is
  suggested.  The proposed method relies on the concept of the ``local
  nuclear character'' of the SRC.  We quantify the SRC and its mass
  dependence by computing the number of independent-particle model
  (IPM) nucleon pairs in a zero relative orbital momentum state.  We
  find that the relative probability per nucleon for 2N SRC follows a
  power law as a function of the mass number $A$. The predictions are
  connected to measurements which provide access to the mass
  dependence of SRC.  First, the ratio of the inclusive inelastic
  electron scattering cross sections of nuclei to $^{2}$H at large
  values of the Bjorken variable.  Second, the EMC effect, for which we
  find a linear relationship between its magnitude and the predicted
  number of SRC-prone pairs.
\end{abstract}
%\end{start}

\maketitle
\section{Introduction}
In a mean-field model fluctuations are completely ignored.  The SRC
induce spatio-temporal fluctuations from the mean-field predictions.
Realistic nuclear wave functions reflect the coexistence of
single-nucleon (mean-field) structures and cluster structures. The
clusters account for beyond mean-field behavior. As the
nucleon-nucleon interaction is short ranged, the clusters attributed
to SRC are predominantly of the two-nucleon (2N) type. Given an
arbitrary nucleus $A(N,Z)$ we address the issue of quantifying the
number of SRC-prone pairs. Our suggested method, albeit approximate,
is robust, model independent, and is applicable to any nucleus from He
to Pb.  Our goal is to come with a systematic insight into the mass
and isospin dependence of the SRC without combining results from
various types of calculations.

\section{Quantifying nuclear correlations}
\subsection{Mean-field approximation and beyond}
\label{sec:IIB}
A time-honored method to account for the effect of correlations in classical
  and quantum systems is the introduction of correlation functions.
  Realistic nuclear wave functions $\mid { \Psi} _A \rangle $ can be
  computed after applying a many-body correlation operator to a Slater
  determinant $\mid \Psi ^{MF} _A  \rangle$
\begin{equation}   
 \mid { \Psi_A}   \rangle =  \frac{1}
{ \sqrt{\langle \ \Psi  ^{MF} _A \mid \widehat{\cal
G}^{\dagger} \widehat{\cal G} \mid \Psi  ^{MF} _A \ \rangle}} \ 
\widehat
{ {\cal G}} \mid  \Psi  ^{MF} _A \ \rangle \; .
\label{eq:realwf}
\end{equation}
The nuclear correlation operator $\widehat{\cal G}$ has a complicated
spin, spin-orbit and isospin dependence but is dominated by the
central, tensor and spin correlations \cite{Roth:2010bm}
\begin{eqnarray}
\widehat{\mathcal{G}}  & \approx &  \widehat {{\cal S}}  
\biggl[ \prod _{i<j=1} ^{A} \biggl(
1 - g_c(r_{ij}) + f_{t\tau}(r_{ij}){S_{ij}} \vec{\tau}_i \cdot \vec{\tau}_j \nonumber \\
& + & 
f_{s\tau}(r_{ij}) \vec{\sigma}_i \cdot \vec{\sigma}_j \; \;  \vec{\tau}_i \cdot \vec{\tau}_j 
\biggr) \biggr] \; , 
\end{eqnarray}
where $g_c$, $f_{t\tau}$, $f_{s\tau}$ are the
central, tensor, and spin-isospin correlation function, $ \widehat
{{\cal S}} $ the symmetrization operator, ${S_{ij}}$ the tensor
operator, and $\vec{r}_{ij}= \frac{ \vec{r}_i - \vec{r}_j }{\sqrt{2}}$. 
The operator $ {S_{ij}} $ admixes relative two-nucleon states of
different orbital angular momentum, is operative on triplet spin
states only, and conserves the total angular momentum of the pair.

The effect of the correlation functions on the momentum distributions
can be roughly estimated from their squared Fourier transforms.  The
relative momentum ($\vec{k}_{12} = \frac{ \vec{k}_1 - \vec{k}_2
}{\sqrt{2}}$) dependence of squared Fourier transform of the tensor
correlation $\mid f_{t \tau} \left( k_{12} \right) \mid ^{2}$ is very
similar to the squared $D$-wave component of the deuteron wave
function $\mid\Psi_{D} \left( k_{12} \right)\mid ^{2}$ \cite{Vanhalst:2012}.
The effect of
the tensor correlation function is largest for moderate relative
momenta $ \left( 100 \lesssim k_{12} \lesssim 500 ~ \textrm{MeV} \right)$. For
very large $k_{12}$, the $g_{c}$ is the dominant contribution.
Whereas a large model dependence for the $g_{c}$ is observed, the
$f_{t \tau}$ seems to be much better constrained.

After introducing the wave functions of Eq.~(\ref{eq:realwf}), the
one-body momentum distributions 
can be written as
\begin{eqnarray}
P_{1}  \left( \vec{k} \right) 
& = &
P^{(0)}_1  \left( \vec{k} \right)  
+  P^{(1)}_1  \left( \vec{k} \right)   \; .
%\\
%P_2 \left( \vec{k}_{12}, \vec{P}_{12} \right) & = &
%P^{(0)}_2 \left( \vec{k}_{12}, \vec{P}_{12} \right) +
%P^{(1)}_2 \left( \vec{k}_{12}, \vec{P}_{12} \right) \; .
%\nonumber \\
%& & 
%\int d \vec{k} _ {1} \int d E_{1}   
%P _{1}  \biggl( \vec{k} , E \biggl|  \vec{k}_{1} , E_{1} \biggr) + \ldots  
%\; .
\end{eqnarray}
The $ P^{(0)}_{1}$ is the mean-field part and is fully determined by
the Slater determinant $ \mid \Psi _{A} ^{MF} \ \rangle$.  The SRC
generate a fat momentum tail to the $ P_{1} \left( \vec{k} \right) $.
The high momentum tails to $ n^{(1)}_{1} \left( k \right) = \int
\mathrm{d} \Omega_{k} P_1^{(1)}(\vec{k}) $ have a very
similar momentum dependence for all nuclei, including the deuteron,
which alludes to an universal character of SRC
\cite{Feldmeier:2011qy}. It has been theoretically predicted
\cite{frankfurt88,CiofidegliAtti:1995qe,janssen00} and experimentally
confirmed in semi-exclusive $A(e,e'p)$ measurements
\cite{Iodice:2007mn} that the major fraction of the high-momentum tail
to $n _{1}^{(1)} \left( k \right)$ strength is contained in
very specific parts of the single-nucleon removal energy-momentum
phase space, namely those where the ejected nucleon is part of a pair
with high relative and small c.m. momentum, the so-called ridge in the
spectral function \cite{janssen00}.

\subsection{Quantifying two-nucleon correlations}

Theoretical $^{16}$O$(e,e'pn)$ calculations
\cite{janssen00,ryck00,Barbieri:2004xn} have predicted that the tensor
parts of the SRC are responsible for the fact that the correlated $(e,e'pn)$
strength is typically a factor of 10 bigger than the correlated $(e,e'pp)$
strength.  Calculations indicated that the tensor correlations are
strongest for pn pairs with ``deuteron-like'' $\left| l_{12} =0,
  S=1\right>$ relative states \cite{ryck00,Barbieri:2004xn}.
Recently, this dominance of the pn correlations over pp and nn ones has
been experimentally confirmed \cite{Subedi:2008zz,
  PhysRevLett.105.222501}.

Accordingly, a reasonable estimate of the amount of correlated nucleon
pairs in $A(N,Z)$ is provided by the number of pairs in a 
$l_{12}=0$ state.  In order to determine that number for a given set
of single-particle states, one needs a coordinate transformation from
$(\vec{r}_1,\vec{r}_2)$ to $ \left(\vec{r}_{12}=\frac { \vec{r}_{1} -
    \vec{r}_2 } { \sqrt{2} }, \vec{R}_{12} = \frac {\vec{r}_{1} +
    \vec{r}_{2} } { \sqrt{2} } \right)$. For a harmonic oscillator
(HO) Hamiltonian this transformation can be done with the aid of
Moshinsky brackets \cite{BookHOMoshinsky}.
After introducing the spin and isospin degrees-of-freedom, in a
HO basis a normalized and antisymmetrized ($na$) two-nucleon state reads
($\alpha _{i} \equiv (n_i l_i j_i t_i)$)
\begin{eqnarray}
&& 
\left| \alpha_{1}   \alpha _{2} ;  J M \right> _{na} 
  = \frac {\left( 1 - \mathcal{P}_{12} \right)
} { \sqrt { 2 \left( 1 + \delta _ {\alpha_{1} \alpha _{2}} \right)}}
\left| \alpha_{1}   \left( \vec{r} _{1} \right) 
       \alpha _{2}  \left( \vec{r} _{2} \right)  
      ;  J M \right> 
\nonumber \\
&& =  
\sum_X
\braket{ n_{12} l_{12} N_{12} \Lambda_{12} LM_L, SM_S TM_T}{\alpha_1 \alpha_2; J_MJ}
\nonumber \\
&&  \times \left|
\left[ {n_{12} l_{12} \left( \vec{r}_{12} \right)}, 
       {N_{12} \Lambda_{12} \left( \vec{R}_{12} \right) }
\right]LM_L, SM_S, TM_T \right> 
 \; ,
\nonumber \\ & & 
\label{eq:2bodynas}
\end{eqnarray}
where $\sum_X$ sums over the appropriate quantum numbers ($n_{12}$, $l_{12}$,
$N_{12}$, $\Lambda_{12}$, $L$, $M_L$, $S$, $M_S$, $T$, $M_T$), 
$\braket{\ldots}{\ldots}$
is the transformation bracket as defined in Ref.~\cite{Vanhalst:2012} and 
$\mathcal{P}_{12}$ the interchange operator for the spatial, spin,  
and isospin coordinate.
Starting from the Eq.~(\ref{eq:2bodynas}) one can compute in a HO
single-particle basis how much a pair wave function with quantum numbers  
\begin{equation}
\left|\left[
n_{12}l_{12} \left( \vec{r}_{12} \right), N_{12}\Lambda_{12} \left( \vec{R}_{12}
    \right) \right] LM_L, SM_S, TM_T \right> 
\label{eq:pairwavefunction}
\end{equation}
contributes to the sum-rule
\begin{eqnarray}
& & \sum _{J M} \sum _{\alpha _{1}  \le \alpha _{F} ^{N_1}}
\sum _{\alpha _{2}  \le \alpha _F ^{N_2}} \;
  _{na} \left< \alpha_{1} \alpha _{2} ;  J M \right.
 \left| \alpha_{1} \alpha _{2} ;  J M \right> _{na}  
\nonumber \\
& &  =  
  \begin{cases}
   \frac{N(N-1)}{2} & N_1 = N_2 = n \\
   \frac{Z(Z-1)}{2} & N_1 = N_2 = p \\
   N Z & N_1 \not= N_2 
  \end{cases}
   \; .
\label{eq:sumrule}
\end{eqnarray}
This can also be done for any other non-relativistic basis $ \left| n l j m \right> $
of single-particle states in a two-step procedure.  First, a 2N state can
be expressed in a HO basis. Second, the Eq.~(\ref{eq:2bodynas})
can be used to determine the weight of the pair wave functions of
Eq.~(\ref{eq:pairwavefunction}).
\begin{figure}
  \centering
 \includegraphics[width=0.49\textwidth]{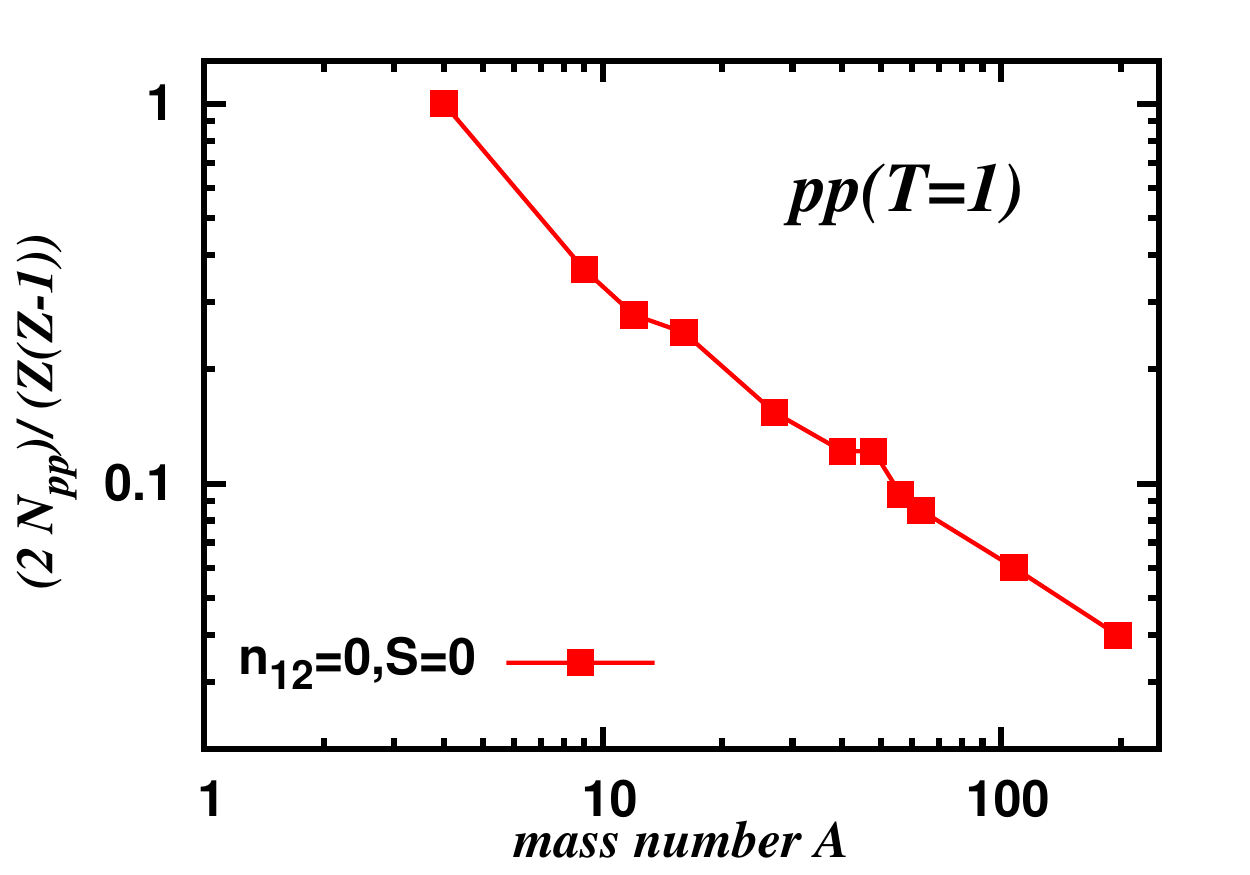}
  \includegraphics[width=0.49\textwidth]{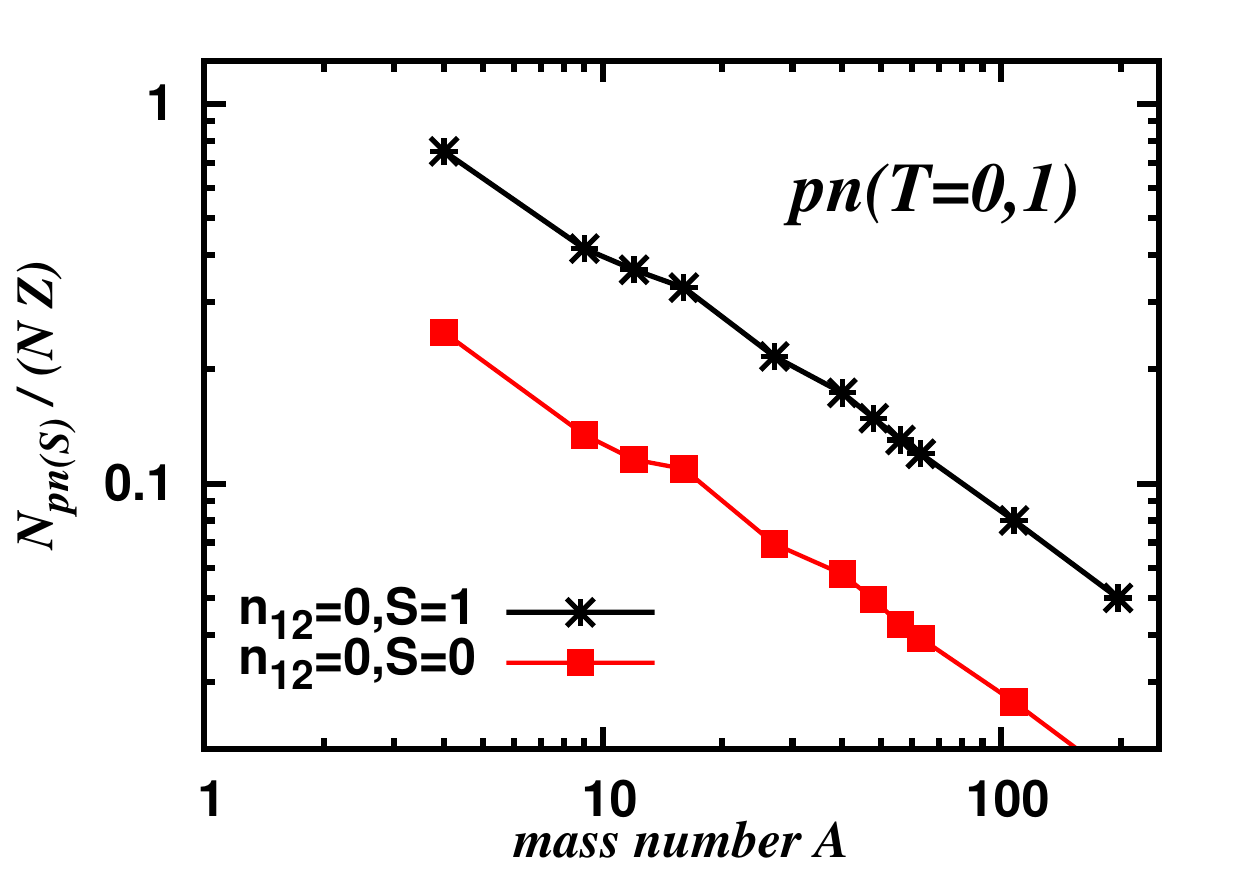}
  \caption{The computed values for $\frac {2} {Z(Z-1)}
    N_{pp} $, and $\frac {1} {(NZ)}
    N_{pn(S)} $ which represent the predicted fraction of the pairs
    which are prone to SRC. The results are obtained for HO
    single-particle wave functions with $\hbar \omega$ (MeV) $= 45 A ^{
      -\frac {1} {3}}- 25 A ^{ -\frac {2} {3}} $ and for the target
    nuclei $^{4}$He, $^9$Be, $ ^{12}$C, $ ^{16}$O, $ ^{27}$Al, $
    ^{40}$Ca, $ ^{48}$Ca, $ ^{56}$Fe, $^{63}$Cu, $ ^{108}$Ag, and $
    ^{197}$Au. The computed values for $nn$ correlations %j$\frac{2}{N(N-1)} N_{nn}$ which
    are similar to the pp results and can and be found in Ref.~\cite{Vanhalst:2012}.
    %**Je zou nn kunnen weglaten en vermelden dat het eruit ziet als PP**
    }
\label{fig:spairnumbers}
\end{figure}
The number of ($n_{12}=0$, $l_{12}=0$) pairs, which are stated to be a
reasonable estimate for the number of the SRC-prone is given by the
expression,
%We used 
%In Fig.~\ref{fig:spairnumbers} we calculated 
%The IPM pp pairs are mainly subject to the central SRC
%which requires them to be close.  
%This implies that a reasonable estimate of the number of IPM pp pairs which
%receive substantial corrections from the SRC is given by an expression
%of the type
\begin{equation}
N_{pp} (A,Z)  =  
\sum _{J M} \sum _{\alpha _{1}  \le \alpha _{F} ^{p}}
\sum _{\alpha _{2}  \le \alpha _F ^p}
{}_{na} \left< \alpha_{1} \alpha _{2} ;  J M \right|
\mathcal{P}_{\vec{r}_{12}}^{n_{12}=0 l_{12}=0}
 \left| \alpha_{1} \alpha _{2} ;  J M \right> _{na}\; , 
\label{eq:project2NSRC}
\end{equation}
where $ \mathcal{P}_{\vec{r}_{12}}^{n_{12}=0 l_{12}=0} $ is a projection
operator for 2N relative states with $n_{12}=0$, $l_{12}=0$.  A similar
expression to Eq.~(\ref{eq:project2NSRC}) holds for the nn pairs. For
the pn pairs it is important to add the projection operator
$\mathcal{P}_{\vec{\sigma}}^{S}$ to discriminate between the triplet
and singlet spin states.  In Fig.~\ref{fig:spairnumbers} we display
some computed results for the $N_{pp}$ and $N_{pn(S)}$ for 11 nuclei
covering the full mass table. Naively one could expect that the number 
of correlated pn (pp) pairs in a nuclei scales like $NZ$ $\left( 
\frac{Z(Z-1)}{2} \right) \sim A^2$. From Fig.~\ref{fig:spairnumbers}
it is clear that the mass dependence of the number of SRC pairs 
approximately follows a power law.
%The pn($l_{12}=0$, $S=1$) 
%pairs scale as $\sim A^{1.46 \pm 0.02}$, the pn($n_{12}=0$,$l_{12}=0$,$S=1$) pairs
%scale as $\sim A^{1.35 \pm 0.03}$.
The pn ($n_{12}=0$, $l_{12}=0$, $S=1$) SRC-prone pairs scale as $\sim A^{1.35 \pm 0.03}$.

The results for $N_{nn}$ which are similar to $N_{pp}$ can be found in 
Ref.~\cite{Vanhalst:2012}.  In Ref.~\cite{Vanhalst:2012} a method to 
estimate the number of
correlated $3N$ clusters is developed.  We find that there is (as for
2N correlations) a power law relation between
the mass $A$ and the number of correlated ppn triples.

\section{Results}
\subsection{Two-body correlations}
\label{sec:resultsB}
Following the experimental observation \cite{PhysRevC.48.2451,
  PhysRevLett.96.082501,PhysRevLett.108.092502} that the ratio of the inclusive
electron scattering cross sections from a target nucleus $A$ and from
the deuteron $D$
\begin{equation}
\frac
{\sigma ^{A} \left(x_{B} , Q ^{2} \right) }
{\sigma ^{D} \left(x_{B} , Q ^{2} \right) } \; ,
\label{eq:defofratiodata}
\end{equation}
scales for $1.5 \lesssim x_{B} \lesssim 2$ and moderate $Q^{2}$, 
it has been suggested \cite{PhysRevC.48.2451} to parameterize the
ratio $\sigma ^{A}/\sigma^{D} $ in the following form
\begin{equation}
a_2 \left(  {A}/ {D} \right) 
= \frac {2} {A} \frac 
{\sigma ^{A} \left( x_{B}, Q ^{2} \right)} 
{\sigma ^{D} \left( x_{B}, Q ^{2} \right)} \; \; \left(1.5 \lesssim x_{B} \lesssim 2 \right) \; .
\label{eq:a2}
\end{equation}

In a simplified reaction-model picture, which ignores for example 
the effect of c.m. motion of pairs in finite nuclei, the quantity $\frac
{A} {2} a_{2} \left( {A} / {D} \right)$ can be connected with the
number of correlated pairs in the nucleus $A$ \cite{PhysRevLett.96.082501}. 
Assuming that all pn
pairs contribute one would expect that for the relative amount of
correlated two-nucleon clusters $a_{2} \left( {A} / {D} \right) \sim
A$.  
We suggest in Refs.~\cite{Vanhalst:2011es,Vanhalst:2012} that the correlated pn
pairs contributing to the $a_2(A/D)$, are predominantly $(T=0, S=1)$
pairs and that $ a_2 (A/D)$ is proportional to 
the per nucleon probability for a pn SRC relative to the deuterium.  
Thereby, the per nucleon probability for a pn SRC
relative to the deuterium can be defined as
\begin{equation}
\frac {2} {N+Z} \frac {N_{pn(S=1)}(A,Z)}{N_{pn(S=1)}(A=2,Z=1)} = 
\frac {2} {A}  {N_{pn(S=1)}(A,Z)} 
\; .
\label{eq:pernucleonSRCrelativetod}
\end{equation}

\begin{figure}
 \centering
 \includegraphics[width=0.7\columnwidth]{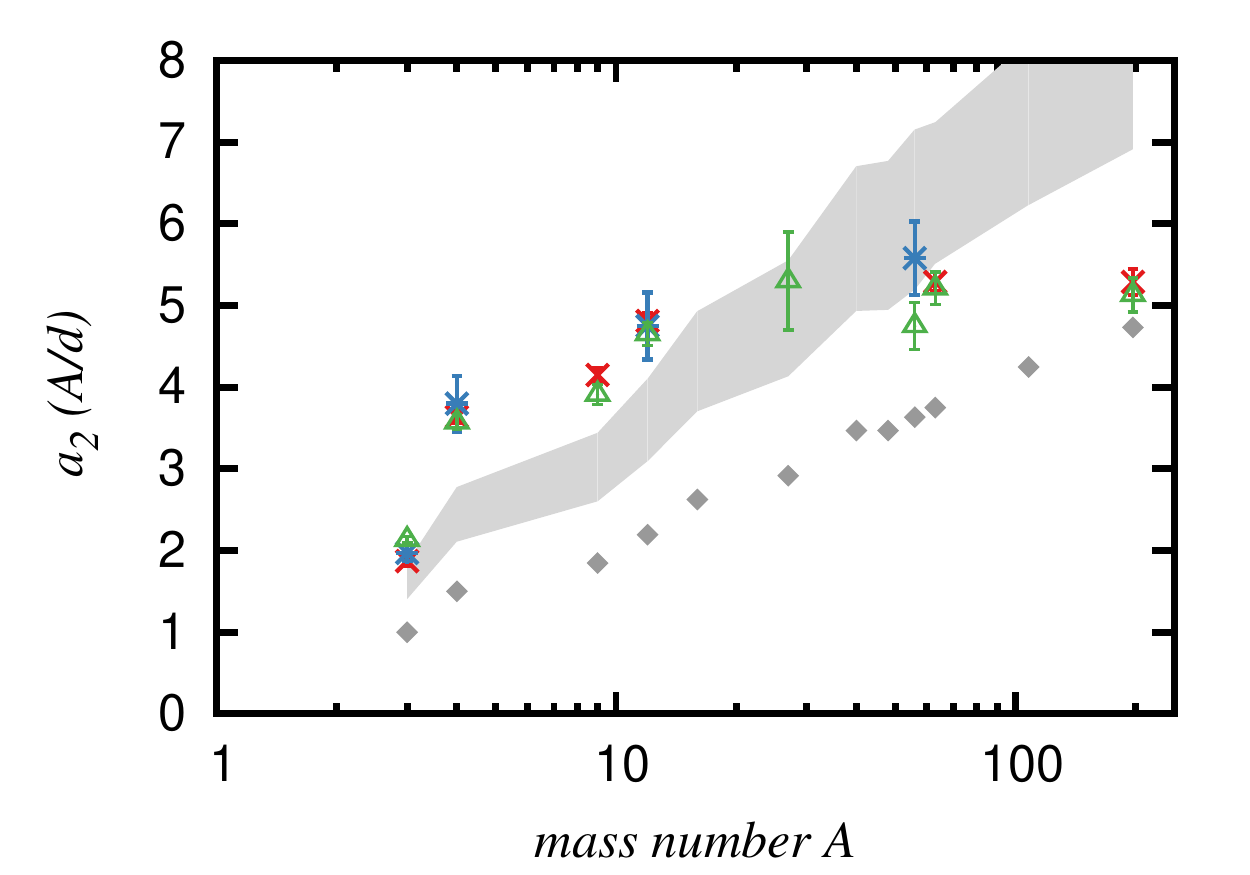}
 \caption{The computed values for the $a_2 (A/D)$ for various
   nuclei. The data are from
   Refs.~\cite{PhysRevLett.96.082501,PhysRevLett.108.092502,Arrington:2012ax}.
   The shaded region is the prediction after correcting the computed
   values of $a_2(A/D)$ for the c.m. motion of the pair. The
   correction factor are determined by linear interpolation of the
   factors listed in Table~1 of Ref.~\cite{Vanhalst:2012}.  The width
   of the shaded area is determined by the estimated errors of the
   c.m. correction factors.}
 \label{fig:a2results}
\end{figure}

Apart from corrections stemming from final-state interactions, a
correction factor which accounts for the c.m. motion of the correlated
pairs blurs the connection between the measured $a_2(A/D)$
coefficients and the number of correlated pairs.  We have opted to
correct the predicted $a_{2}$ coefficients and not the data for
c.m. motion.  The magnitude of the c.m. motion correction factor is
subject of ongoing discussions \cite{Arrington:2012ax} and is far 
from established.  We stress that the c.m. correction factor cannot 
be computed in a model-independent fashion. To estimate
the c.m. correction factor, we have simulated the number of events 
in the probed phase with and without accounting for pair c.m. motion.
We simulate the interaction of a virtual photon with a nucleon pair
inside a nucleus. We assume that the virtual photon reacts instantly
with one of the nucleons inside the pair, i.e. the virtual photon
is entirely absorbed by one of the paired nucleons. 
In Ref.~\cite{Vanhalst:2012} we stated  a c.m.  correction factor of
$1.7\pm0.3$ which shows little mass dependence.  
For light nuclei our predictions corrected for c.m. motion of the
pairs, underestimate the measured $a_{2}$.  This may be attributed
to the lack of long-range clustering effects in the adopted wave
functions. Indeed, it was pointed out in
Ref.~\cite{PhysRevC.83.035202} that the high-density cluster
components in the wave functions are an important source of
correlation effects beyond the mean-field approach.  For heavy nuclei
our predictions for the relative SRC probability per nucleon do not
saturate as much as the data seem to indicate.  We stress that
final-state interactions (FSI) represent another source of corrections
which may induce an additional $A$-dependent correction to the data.
FSI of the outgoing nucleons with the residual spectator nucleons,
could shift part of the signal's strength out (or, in) of the cuts
applied to the experimental phase space and decrease (or increase) the
measured cross section and the corresponding $a_2$ coefficient.

\subsection{EMC effect}

In 1983, the EMC collaboration discovered that the ratio $R(x_B)$ of
the Deep Inelastic Scattering (DIS) cross section of leptons on a
nucleus and the deuteron differs from one \cite{Aubert:1983rq}.  
At medium Bjorken $x_B$-values, $0.3 \leq x_B \leq 0.7$, $R(x_B)$ drops from
approximately one to values as low as $0.8$.  This effect is known as
the EMC effect. This reduction of $R(x_B)$ is not easily explained and
so far there still is no established explanation yet.  More recently,
a linear relation between the slope of the EMC effect
$-\frac{\mathrm{d} R}{ \mathrm{d} x_B}$ in the region $0.3 \leq x_B
\leq 0.7$, and the SRC scaling factor $a_2(A/D)$, obtained from
inclusive electron scattering, has been found \cite{Weinstein:2010rt}.
Consequently, one may expect that $ - \frac{ \mathrm{d} R}{
  \mathrm{d}x_B}$ could be related to the number of SRC-prone pairs in the
nucleus.  When computing the $a_2(A/D)$ coefficients we included the
SRC-prone $\left( S=1, T=0 \right)$ pn pairs. This is justified by
the dominance of the tensor correlation in the inclusive electron
scattering data at moderate momentum transfers and high $x_B$.
In the DIS experiments, which are performed at considerably
higher $Q^2$, partons are the relevant degrees of freedom and one 
may argue that all correlated 2N pairs should be counted equally.
Therefore when relating
$-\frac{\mathrm{d}R}{\mathrm{d}x_B}$ to the number of correlated
pairs, one should count all SRC-prone pairs including the $\left(
S=0, T=1 \right)$ pairs.

\begin{figure}
  \centering
\includegraphics[angle=-90,width=0.7\columnwidth]{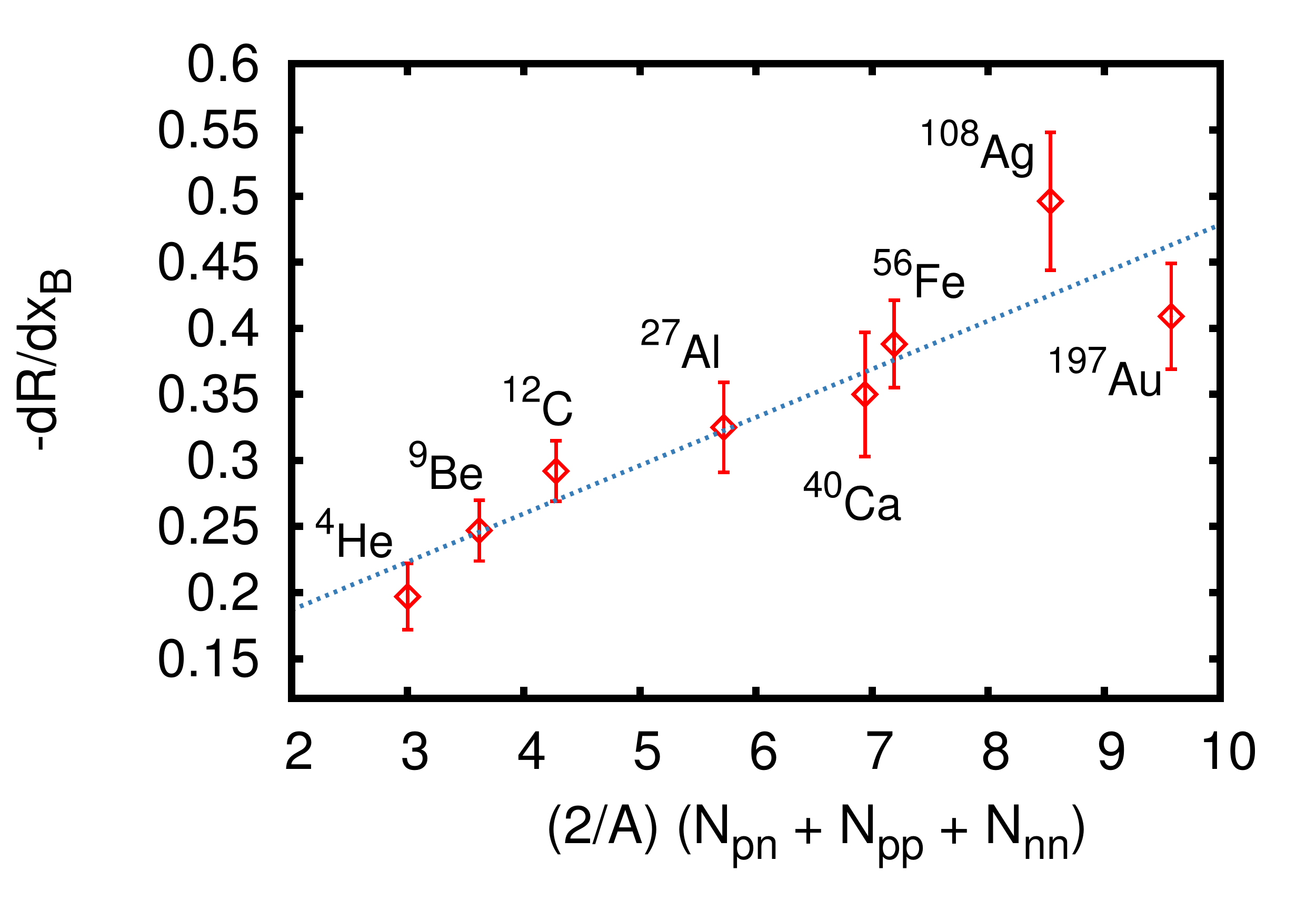}
\caption{The magnitude of the EMC effect versus our predictions for the
``per nucleon probability for 2N SRC relative to the deuteron'',
$\frac{2}{A} (N_{pn(S=1)} + N_{pn(S=0)} + N_{pp} + N_{nn} )$.
 The data are from the analyses presented in
  Refs.~\cite{Seely:2009gt,PhysRevD.49.4348,Arrington:2012ax}. The
  fitted line obeys the equation $-\frac{d R}{d x_{B}} = (0.11 \pm
  0.03) + (0.036 \pm 0.005) \cdot \frac {2} {A} (N_{pn}+N_{pp}+N_{nn}) $. }
\label{fig:EMCeffect}
\end{figure}

\begin{figure}
  \centering
\includegraphics[angle=-90,width=0.7\columnwidth]{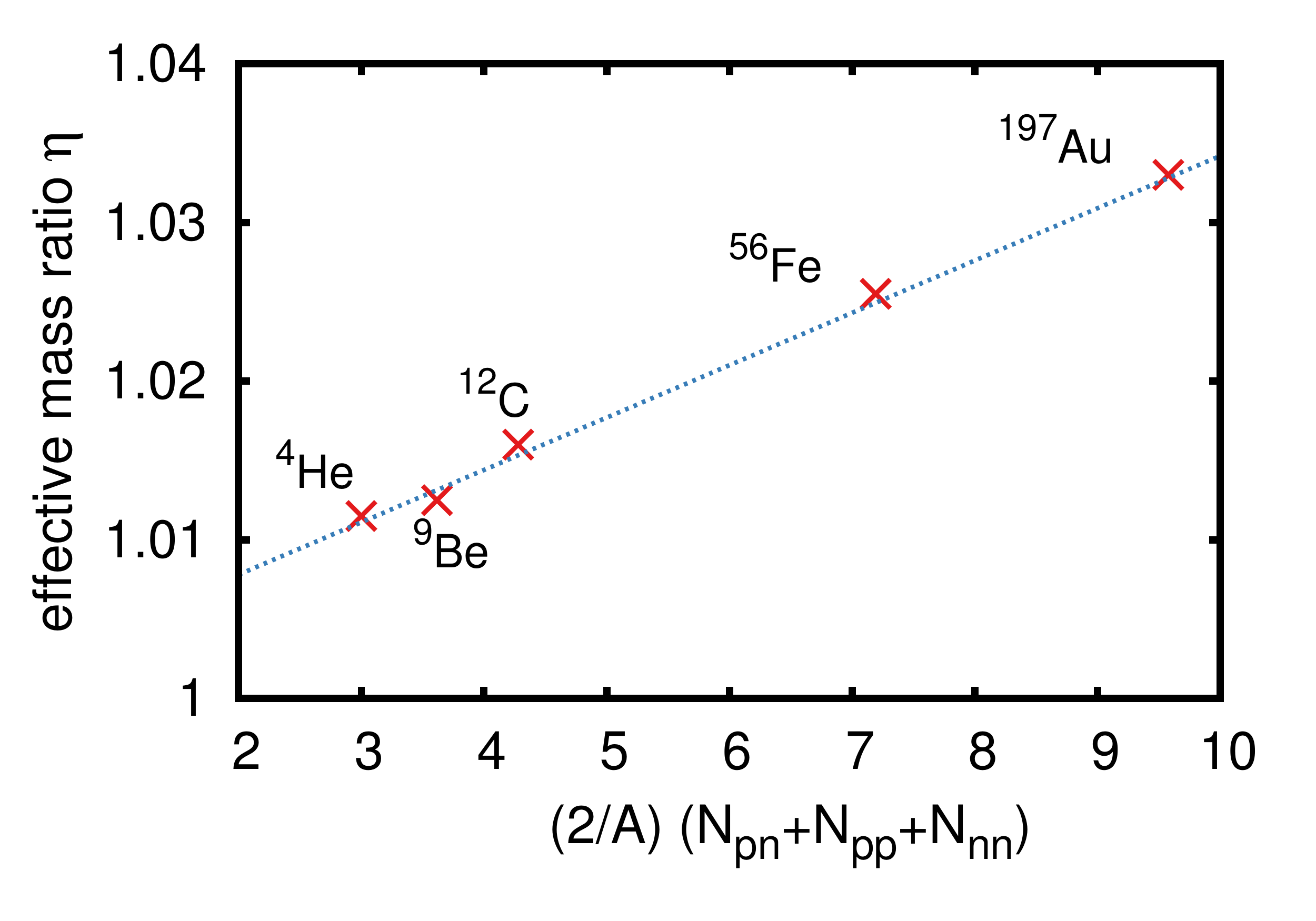}
\caption{ Relation between the
  effective mass ratio $\eta=\frac{m}{m^*}$ of Ref.~\cite{canal2012emc} and the
  ``per nucleon probability for 2N SRC relative to deuteron''. The
  fitted line obeys the equation $\eta= (1.0012 \pm 0.0006) + (0.0033
  \pm 0.0001) \cdot \frac{2}{A}(N_{pn}+N_{pp}+N_{nn})$.}
\label{fig:eff_mass}
\end{figure}

In Fig.~\ref{fig:EMCeffect} we display the magnitude of the EMC
effect, quantified by means of $- \frac {d R} {d x_{B}}$ versus our
predictions for the ``per nucleon probability for 2N SRC relative to the deuteron'',
or $\frac{2}{A} (N_{pn(S=1)} + N_{pn(S=0)} + N_{pp} + N_{nn} )$.
We stress
that the numbers which one finds on the x-axis are the results of
parameter-free calculations. We consider the "per nucleon probability
for 2N SRC relative to the deuteron" as a measure for the magnitude of
the nucleon-nucleon SRC in a given nucleus. As can be seen in
Fig.~\ref{fig:EMCeffect}, there is a nice
linear relationship between the quantity which we propose as a per
nucleon measure for the magnitude of the SRC and the magnitude of the
EMC effect. 

In Ref.~\cite{PhysRevLett.53.1430} a formalism to describe the EMC effect was developed by introducing 
an effective mass.
In this formalism the nucleons bound in a nuclei are assumed to have a different 
effective mass $m^*$ than the free nucleon mass $m$.
A calculated ratio of nuclear to free structure function is fitted to the 
experimental values of the nucleus to deuteron structure function.
The nuclear structure function is a convolution of the free structure function
and a distribution function which accounts for Fermi smearing and binding effects.
The effective mass remains the only free parameter which is fixed by the fit.
The formalism is quite efficient in describing the EMC data.
In Fig.~\ref{fig:eff_mass} we relate the latest value for the ratio of the free nucleon mass
to the effective one, $\eta= \frac{m}{m^*}$ \cite{canal2012emc},
to our calculated ``per nucleon probability for 2N SRC relative to deuteron''.
It is obvious that there is a linear relation between the effective mass parameter,
used to describe the EMC effect and our ``per nucleon probability for 2N SRC relative
to the deuteron'', which is our estimate for the amount of nucleon-nucleon pairs in a
nucleus prone to correlations.

\section{Conclusion}
We have provided arguments that the mass dependence of the magnitude
of the NN correlations can be captured by some approximate
principles.  Our method is based on the assumption that correlation
operators generate the correlated part of the nuclear wave function
from that part of the mean-field wave function where two nucleons are
spatially ``sufficiently close''.
We have calculated the number of pn, pp and nn ($n_{12}=0$, $l_{12}=0$) SRC-prone
pairs and studied their mass and isospin dependence.  The $A$ dependence of
the magnitude of the pp, nn, and pn SRC can be captured in a in a
power-law dependence, $A^{\alpha}$ with $\alpha=1.35 \pm 0.03$.

We related the experimentally determined scaling parameter $a_2\left(A/D\right)$
to our computed "per nucleon probability for a pn SRC relative
to deuterium``.
To connect the
computed number of SRC pairs to the measured $a_2\left(A/D\right)$ corrections are in order. Published
experimental data include the radiation and Coulomb corrections.  The
correction factor stemming from final-state interactions and from the
c.m. motion of the correlated pair, however, is far from established.
After correcting for the c.m. motion of pairs in a finite nuclei,
our model calculations for $a_2$ are of the right order of magnitude.
We predict a rather soft mass dependence which for heavy nuclei,
however, is stronger than what the experiments indicate. It remains
to be studied whether final-state interactions can account for this
additional mass dependent correction factor.
%and capture the $A$-dependence qualitatively.

We find a
linear relationship between the magnitude of the EMC effect and the
computed per nucleon number of SRC-prone pairs.
Also other parameters used to describe the EMC effect, like the 
effective mass parameter, tend to have a linear relationship to 
our predictions for the per nucleon number of SRC 2N pairs.
Those may indicate that
the EMC effect is (partly) driven by local nuclear dynamics
(fluctuations in the nuclear densities), and that the number of SRC-prone 
pairs serves as a measure for the magnitude of this effect.
 
\subsection*{Acknowledgments}
The computational resources (Stevin Supercomputer Infrastructure) and
services used in this work were provided by Ghent University, the
Hercules Foundation and the Flemish Government – department EWI.  This
work is supported by the Research Foundation Flanders.


\begin{thebibliography}{99}
%\bibitem{myref1} A.V. Mikhailov, \textit{Physica D} \textbf{3} (1986) 1872-1873.
%\bibitem{myref6} P. Ring and P. Schuk, \textit{The Nuclear Many-Body Problem}, Springer, Berlin (1980).
%\bibitem{myref18} J. Doe,  ``Title of Paper,'' \textit{Name of Journal} (to be published).
\bibitem{Roth:2010bm} R. Roth, T. Neff and H. Feldmeier, \textit{Prog. Part. Nucl. Phys.} \textbf{65} (2010) 50.
\bibitem{Vanhalst:2012} M. Vanhalst, W. Cosyn and J. Ryckebusch, \textit{Phys. Rev. C} (to be published) (2012) nucl-th/1206.5151.
\bibitem{Feldmeier:2011qy} H. Feldmeier, W. Horiuchi, W. Neff and Y. Suzuki, \textit{Phys. Rev. C} \textbf{84} (2011) 054003.
\bibitem{CiofidegliAtti:1995qe} C. Ciofi degli Atti and S. Simula, \textit{Phys. Rev. C} \textbf{53} (1996) 1689.
\bibitem{janssen00} S. Janssen, J. Ryckebusch, W. Van Nespen and D. Debruyne, \textit{Nucl. Phys. A} \textbf{672} (2000) 285.
\bibitem{frankfurt88} L. Frankfurt and M. Strikman, \textit{Phys. Rep.} \textbf{160} (1988) 235.
\bibitem{Iodice:2007mn} M. Iodice, E. Cisbani \etal, \textit{Phys. Lett.} \textbf{B653} (2007) 392.
\bibitem{Barbieri:2004xn} C. Barbieri, C. Giusti, F. Pacati and W. Dickhoff, \textit{Phys. Rev. C} \textbf{70} (2004) 014606.
\bibitem{ryck00} J. Ryckebusch, S. Janssen, W. Van Nespen and D. Debruyne, \textit{Phys. Rev. C} \textbf{61} (2000) 021603R.
\bibitem{Subedi:2008zz} R. Subedi, R. Shneor, P. Monaghan, B.D. Anderson \etal, \textit{Science} \textbf{320} (2008) 1476.
\bibitem{PhysRevLett.105.222501} H. Baghdasaryan \etal\; (CLAS Collaboration), \textit{Phys. Rev. Lett.} \textbf{105} (2010) 222501.
\bibitem{BookHOMoshinsky} M. Moshinsky and Y.F. Smirnov, \textit{The harmonic oscillator in modern physics}, Harwood Academic Publischers, Amsterdam (1996).
\bibitem{PhysRevC.48.2451} L. Frankfurt, M. Strikman, D. Day and M. Sargsyan, \textit{Phys. Rev. C} \textbf{48} (1993) 2451.
\bibitem{PhysRevLett.96.082501} K.S. Egiyan \etal\; (CLAS Collaboration), \textit{Phys. Rev. Lett.} \textbf{96} (2006) 082501.
\bibitem{PhysRevLett.108.092502} N. Fomin, J. Arrington \etal, \textit{Phys. Rev. Lett.} \textbf{108} (2012) 092502.
\bibitem{Arrington:2012ax} J. Arrington, A. Daniel, D. Day, N. Fomin, D. Gaskell \etal, \textit{arXiv} (2012) nucl-ex/1206.6343.
\bibitem{Vanhalst:2011es} M. Vanhalst, W. Cosyn and J. Ryckebush, \textit{Phys. Rev. C} \textbf{84} (2011) 031302.
\bibitem{PhysRevC.83.035202} M. Hirai, S. Kumano, K. Saito and T. Watanabe, \textit{Phys. Rev. C} \textbf{83} (2011) 035202.
\bibitem{Aubert:1983rq} J.J. Aubert \etal\; (European Muon Collaboration), \textit{Phys. Lett. B} \textbf{123} (1983) 275.
\bibitem{Weinstein:2010rt} L.B. Weinstein, E. Piasetsky, D.W. Higinbotham, J. Gomez, O. Hen and R. Shneor, \textit{Phys. Rev. Lett.} \textbf{106} (2011) 052301.
\bibitem{Seely:2009gt} J. Seely, A. Daniel, D. Gaskell, J. Arrington, N. Fomin \etal, \textit{Phys. Rev. Lett.} \textbf{103} (2009) 202301.
\bibitem{PhysRevD.49.4348} J. Gomes, R. Arnold, P. Bosted, C. Chang \etal, \textit{Phys. Rev. D} \textbf{49} (1994) 4348.
\bibitem{PhysRevLett.53.1430} C.A. Garcia Canal, E.M. Santangelo and H. Vucetich, \textit{Phys. Rev. Lett.} \textbf{53} (1984) 1430.
\bibitem{canal2012emc} C.A. Garcia Canal, T. Tarutina and V. Vento, \textit{arXiv} (2012) hep-ph/1206.1541v2.
\end{thebibliography}
\end{document}